# An Exploration of Effects of Dark Mode on University Students: A Human Computer Interface Analysis

*Awan Shrestha*

*Department of Computer Science and Engineering, Kathmandu University, [awanshrestha1@gmail.com](awanshrestha1@gmail.com)*

*Sabil Shrestha*

*Department of Computer Science and Engineering, Kathmandu University, shresthasabil62@gmail.com*

*Biplov Paneru*

*Department of Electronics and Communication Engineering, Pokhara University, Nepal*

*Biplovp019402@nec.edu.np*

*Bishwash Paneru*

*Department of Applied Science and Chemical Engineering, Pulchowk Engineering Campus, Tribhuvan University, Nepal*

*rampaneru420@gmail.com*

*Sansrit Paudel*

*Department of Computer Science and Engineering, Kathmandu University, 2sansrit@gmail.com*

*Ashish Adhikari*

*Department of Computer Science and Engineering, Kathmandu University, [amazingashish061@gmail.com](amazingashish061@gmail.com)*

*Sanjog Chhetri Sapkota*

*Nepal Research and Collaboration Center, kshetrisanjog@gmail.com*

Corresponding author email: biplovp019402@nec.edu.np

**Abstract**

This research dives into exploring the dark mode effects on students of a university. Research is carried out implementing the dark mode in e-Learning sites and its impact on behavior of the users. Students are spending more time in front of the screen for their studies especially after the pandemic. The blue light from the screen during late hours affects circadian rhythm of the body which negatively impacts the health of humans including eye strain and headache. The difficulty that students faced during the time of interacting with various e-Learning sites especially during late hours was analyzed using different techniques of HCI like survey, interview, evaluation methods and principles of design. Dark mode is an option which creates a pseudo inverted adaptable interface by changing brighter elements of UI into a dim-lit friendly environment. It is said that using dark mode will lessen the amount of blue light emitted and benefit students who suffer from eye strain. Students' interactions with dark mode were investigated using a survey, and an e-learning site with a dark mode theme was created. Based on the students' comments, researchers looked into the effects of dark mode on HCI in e-learning sites. The findings indicate that students have a clear preference for dark mode: 79.7% of survey participants preferred dark mode on their phones, and 61.7% said they would be interested in seeing this feature added to e-learning websites.

## 1 . INTRODUCTION

The time spent by students using digital devices for learning is increasing rapidly with the development of new, portable and instantly accessible technology, such as smartphones and digital tablets. Furthermore, during the pandemic, online learning has dramatically increased. This explosion in e-Learning which is traditionally different from pen and paper has brought a difficult situation because screen and paper have different natures. Although the screen has some advantages over paper, the screen glows and emits blue light to show the content, while paper doesn't. The emission of bright lights, especially blue light during night from a screen has many negative consequences for a user in the long term. [1] While studying online, students have to view the screen for a longer period of time. The brightness of the screen can be a bottleneck for students to study for a longer period of time. So, to reduce the emission of blue light, dark mode is a simple but a great option to have. [2]

Implementing dark mode is not just about making the background darker, but also about balancing the color combination, contrast between background and items such as icons, text, images and many more UI elements. [3] Dark mode provides a great relief on minimizing the dry and painful eyes syndrome that are caused due to studying on a bright blue screen all night long.

Bright lights also affect secretion of melatonin, a hormone needed for sleep. [4] Besides the health benefits, dark mode on various applications helps to enhance battery life. Using dark mode on an OLED screen can extend battery life in noticeable figures. [5] The dark mode created is analyzed using principles of HCI. Human Computer Interaction (HCI) is understood as the design, evaluation and implementation of interactive computing systems for human use. HCI is a key part of computer science, psychology, and design because designing a successful interface requires an understanding of how people interact with devices, user interactive graphical and collaborative platforms.

Dark mode is a new trend in UI/UX design. Almost all Silicon Valley companies are adopting this trend. [6] Today, almost all the major websites support dark mode. Even the Operating Systems are adopting dark mode, including Windows, IOS and Android, in their latest update. Facebook, Twitter, YouTube, Reddit etc. are few big companies that have already included dark mode in their products. Dark mode even though a simple concept has significant advantages to other interface designs, which is why the trend is still growing, and rapidly being adopted by companies worldwide [7]. Tested in an exploratory project involving apartment selection, a human-computer interface was created in a study by Archer et al. [8] to investigate user preferences and the efficacy of output styles and levels of information abstraction in a decision-making environment. The study discovered that voice alone was not as effective for information search as text plus voice. Nevertheless, there were no appreciable differences in preferences between text and voice or text and text plus voice, and task performance was more influenced by cognitive style than by task domain experience. The first online brain-computer interface using deep learning and a menu-driven language generating system based on referencing expressions were created in a study by Kuhner et al. [9] for a BCI-controlled autonomous robotic service assistant. To determine the efficacy of this system, it was connected with a modular ROS-based mobile robot and tested experimentally on a real robot. To investigate user preferences, the efficacy of output modalities, and the degrees of information abstraction in decision making, a human-computer interface was created in a study by Yang et al. [10]. The exploratory study on apartment selection identified disparities between perceived relevance and actual utilization of information qualities and abstraction levels, showing that text with voice was favored over voice alone and that text was the most effective for information search. Software designers utilize anthropomorphism to improve interface usability as computers get more sophisticated. In the study by Laere et al. [11], participants' reflected assessments and self-appraisals were strongly impacted by computer input, but there was no discernible difference in the impact between the two interface types. The study examined human-like and machine-like interaction styles. Software designers utilize anthropomorphism to improve interface usability as computers get more sophisticated. In the study [12] by Mukhopadhyay et al., participants' reflected assessments and self-appraisals were strongly impacted by computer input, but there was no discernible difference in the impact between the two interface types. The study examined human-like and machine-like interaction styles.

The use of eye movements in user interfaces for human-computer interaction control and usability study was covered in the chapter by Birbaumer et al. [13]. Eye movements were recorded, analyzed for usability after the fact, and also used in real time as an input method. This demonstrated the promising potential of integrating eye-tracking in human-computer interaction, as it could be used as the only input for users with disabilities or hands-occupied applications, or it could be combined with other inputs. Jaklič et al. in the study [14] discussed video-conferencing systems that were created to enable reliable and effective communication across vast distances utilizing the Internet and personal computers. Eye contact has always been important for this kind of communication. Since cameras are usually mounted above computer monitors, making eye contact can be difficult. This article suggests a straightforward method to improve the perception of eye contact based on how 3D scenes are viewed on slanted surfaces, which is supported by experimental evidence. Despite being inexpensive, these systems have not been widely adopted.

Mobile device interface and interaction design has a big impact on cognitive offloading. Compared to less responsive or mouse-based controls, participants offloaded more information while using highly responsive touch-based controls, and subjective measurements suggested that metacognitive judgments were also involved in this process as discussed in [15] by Grinschl; et al.. According to the study [16] by Zaina et al. Barriers to accessibility may be generated while developing mobile applications, frequently as a result of standard software development UI design patterns. These hurdles must be understood in the context of the challenges faced by software experts; recommendations have been put forward to prevent accessibility concerns in mobile development. Two innovative gaze-based predictive user interfaces that are able to dynamically provide adaptive interventions in line with the user's objectives and intents linked to the task were offered. In addition to managing uncertainty in prediction model outputs, these interfaces preserve usability and do not negatively impact perceived job burden as mentioned in the study [17].

A cross-stitch technique was used to construct a wearable tactile sensor array made of fiber that was incorporated into human-machine interfaces in the study [18] by Zhong et al.. The system, which included sound feedback components, microprocessors, and sensors, allowed for an interactive process. To speed up reaction times and enhance operational behavior, a quick signal prediction technique was suggested. Many social media, websites and applications are introducing the dark mode as an additional feature. Within just a span of a couple of years, almost every popular website, app and even operating systems now have dark mode as their integral feature. The e-learning sites are also adopting this feature. But the thing that is different about the e-learning sites is that they are the replacement of books and copies. While on the other sites, it might be for many different use cases, in e-learning sites, the main purpose is to learn and teach. While this learning teaching methodology has been long practiced with the help of paper books and copies, the introduction of screens in their replacement will surely have impacts on their users and uses. The dark mode replaces the typical dark text on white background view to white text on dark background view.

So, this research is the study about the impacts of dark mode on HCI in students. Our hypothesis is that, for those students who use the computer screen for longer duration for study, dark mode helps to reduce eye strain and motivates the students to study for a longer period of time in comparison to normal light mode. In this paper, we discuss and investigate this topic on an example of the MOOC system which is used in our university as an online learning platform.

## 2  METHODOLOGY

The paper describes the identification of a problem on user interface design, the techniques for evaluation and implementation of various HCI principles for making several modifications to the design in order to enhance the user interface and experience of online learning platforms. A series of sequential and pre-planned methodologies for evaluating the problems and designing methodologies are used.

## 2.1 Questionnaire and Survey

To deep dive into our case and to understand the demographics of dark mode user students, we conducted a general survey targeting university students, where we asked the students about their views and preference on the dark mode. The survey was conducted online using Google Forms to present the survey questionnaires to the users and record the responses. Survey instruments like Likert Scale (from strongly agree to strongly disagree) were used along with Bipolar semantically anchored items.

Some questions were closed ended leading questions to force users to select a specific answer among the given many whereas some of them were multiple choice for users to pick as many as they resonate with. A total of 31 questions were asked (excluding the user's name). The questions strictly focused on the user interfaces of different themes, likes and dislikes for dark mode, effects of light mode and some problems regarding eyes caused by excessive use of digital screens.

- 650 students participated in the general survey which gave us enough insights to move ahead in the research. Then the experiment was conducted to address the research questions. The dark theme of MOOC was developed using HCI principles. And participants were asked to go through the dark mode of site and read the same text until the time they liked.

- 120 students participated in the experiment and the participants were asked the questions to address the research questions.

**In particular, the following research questions were targeted to audience:**
Q1 Are there any subjective or objective benefits of using dark mode for visual fatigue?
Q2 Do students subjectively prefer dark mode or normal mode in their daily use cases?
Q3 Do students tend to study for a longer period in the e-learning site with dark mode?
.

## 2.2 YouTube

The dark mode of YouTube started rolling out from 2017 as a beta version, and it was officially launched in early 2018, with good response throughout the community, and making headlines at that time. YouTube dark mode was one of the most requested reddit divisions. [8]

## 2.3 Facebook

The dark mode in Facebook was initially limited to messenger apps in early 2017. After getting an overwhelmingly good response from users, the dark mode Instagram app for both Android and IOS were launched in the next update. [9] Currently, all of the Facebook products have dark mode enabled, except

Facebook mobile app, which is currently in beta version of dark mode, and might be released in the next update of app.

## 2.4 Twitter

Twitter has started rolling out pitch-dark night mode called Lights Out for users on Android. The feature was earlier available on the Twitter for IOS app in March and now the same makes its way to Android phones. [10] Twitter also contains an option of optimization of the dark mode by the user itself. Twitter has the option of very dark, mild dark and light mode. The buttons are also customizable by choice of user.

## 2.5 Coursera

Coursera is an Online platform which launched it's dark mode on October 18, 2019. This update is targeted for the students who learn at night, in low-light settings. The new dark mode feature allows learners using a device that runs on iOS 13 (the latest iOS mobile operating system) to apply darker color schemes to their Coursera app. iPhones and iPads running iOS 11 or iOS 12 will continue to see the normal light mode. Many learners find dark mode to be more comfortable to view at night or in a dark room. [11]

### 2.5.1 Identification and evaluation of design.

It is a user centered design process that incorporates ethnographic methods for gathering data relevant to the product via field studies, personal development and user environment design. We were able to filter out several issues related to the design and functionality of MOOC websites using techniques of contextual inquiry, interpretation sessions, persona development and storyboarding. A fictional person named Hari Poudel was created who uses MOOC for a longer period of time, and this project was developed from his perspective.

Several paper prototyping and mockup models were made and tested among several fictitious users, recording the likes and dislikes which was again used to refine the model itself. Same users were given the freedom of competitive usability testing from repeated walkthroughs of the re-designed websites enlisting some more issues in design.

### 2.5.2 Participatory Design Method.

This approach involves the targeted users themselves in the design process of the user interface. We were able to engage the active MOOC user in this approach. We simply created a questionnaire in terms of prevalent issues highlighted by rapid contextual designs and the necessary modifications they would make in order to increase user experience and ease of use by eliminating those issues were duly noted. Real time involvement of the users of different groups pitching their idea on the efficiency of the design leading to better user experience were shared to and fro, which would turn out to be leading stairs to improving the overall system design. Also, with the results of the survey we were able to find the appropriate color combinations and the missing feature like user control of freedom in task submission and also recognition rather than recall in dashboard i.e. To Do Section was missing. We also found that the navigational system was a problem during accessing the content of the previous lectures.

## 2.6 Design Principles and Implementation

*2.6.1 Implementation of Design Principles.*

*2.6.1.1 Visibility*

Visibility is the most important design principle in dark mode implementation. Any user interface can be converted into dark mode just by applying dark colors but the interface might not be good visibly. Visual clues in the website are well rationalized, explained, interpreted and memorized. Implementing dark mode is not just about darker colors. It's about the contrast and readability that a user gets after switching into a dark mode. The interface should make the user's experience more delightful. The e-Learning platforms which only functions in light mode converted to dark mode by just simply changing the background color to dark creates hard text Visibility, low Contrast of icons, text, menu and images, color combinations. These problems identified during evaluations were properly noted and then re-designed keeping visibility in mind. Color contrast, visibility of text, icons were properly selected compared with the background color.

*2.6.1.2 Feedback*

Feedback is necessary for any platform to update users about the condition of the system after the input. Sending information back to the user about what has been done is very important. In Dark mode normal feedback color such as pure red for alert is difficult to see. This kind of problem in feedback is eliminated by using a brighter color which is in contrast with surrounding UI.

*2.6.1.3 Mapping*

Mapping correlates the design of icons, elements of artifacts to the real world. Users must be able to correlate, and use the interface more easily if real world symbols are used for designing interfaces. In Dark mode due to the dark background the icons look saturated, forcing users for more memory load. This was avoided by applying same dark icons.

*2.6.1.4 Consistency*

Consistent sequences of actions should be required in designing the basic UI of a website, The UI in dark mode is consistent with its counterpart in light mode. Even though few changes are seen in dark mode, the overall interface is consistent to light mode.

*2.6.1.5 Affordance*

Affordance of any design refers to an attribute of an object of that design that allows people to know how to use it. Designing products with affordances in mind is vital for clearly showing what users can do with them so dealing with the probable design options were based on the correlation of everyday items and choosing those options with its design and its functionality.

*2.6.2 Implementation of HCI Guidelines.*

Darker modes or environments in e learning platform is intended to reduce blue light and over brightness brought by white and lighter background of any forms during late hours which results in several health conditions, decreased performance, and obstructs a good user experience. So, we focused on keeping dark mode close to the original design in light mode. The contrast and readability that a user gets after switching into a dark mode was created with limited impact on overall design features.

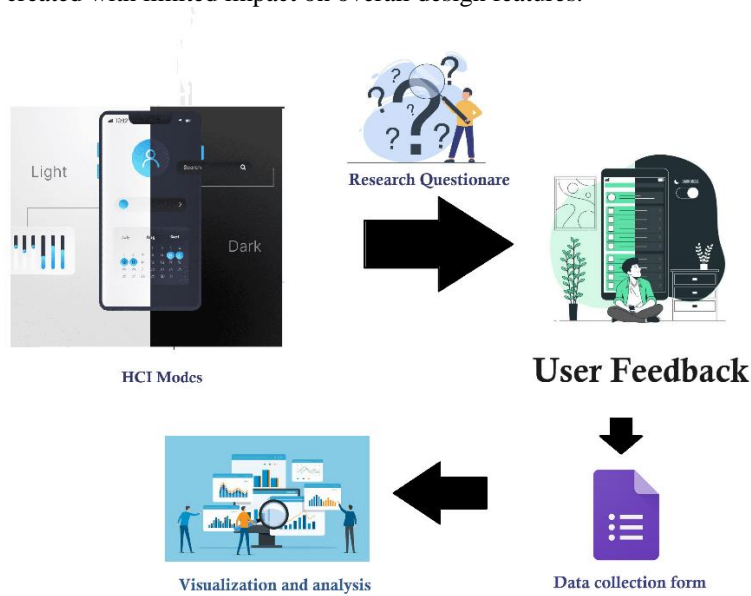

Fig 1. Proposed workflow

The user's skill spectrum can be divided into three parts: beginners, intermediate and experts. Beginners are the users with experience of using dark mode, intermediate users have little knowledge about the interface but can sometimes struggle. Expert users are more proficient in getting interface easily. In reference to these variety of user's skill Levels, toggle between a dark mode and light mode is done by a button which is permanently fixed to the top right corner of the interface. The toggle button has two symbols: moon and sun. Moon representing dark mode and the sun representing light mode. This method makes the toggle button visible to all types. The data is to be collected using forms submission survey as seen in figure 1.

*2.6.3 Implementation of Jackob Neilson.*

*2.6.3.1 Visibility of system status*

The new MOOC in dark mode has a clear visibility even after the change in new color system. The dark mode gives the pleasant effect to the eyes. The color combinations are very beautiful, and all the courses of the weeks are provided with the dropdown making the visibility even clearer.

*2.6.3.2 Match between system and real world*

The new dark mode design's controls follow real-world conventions and correspond to desired outcomes. As in the real world we turn on the light when it gets dark, similarly we can toggle the dark mode in dim light environment to reduce the strain due to bright white background of the screen. And the sun moon icon in the toggler makes the system match with the real world.

*2.6.3.3 User control and freedom*

The new design provides the User control of Freedom. For example, in the assignment section we can clearly view the task, un-submit the task and again resubmit the task. This feature provides users with control of the tasks and prevents error.

*2.6.3.4 Consistency and standards*

The dark theme maintains the dark look and feel throughout the webpage with pleasant look maintaining the standards. Also, the icon colors have been changed accordingly to maintain the consistency.

*2.6.3.5 Error prevention*

In the new system as seen in figure 2(a), there is option to un-submit the task in case of any error in the submission and user can also see the submitted files which helps users to prevent the errors.

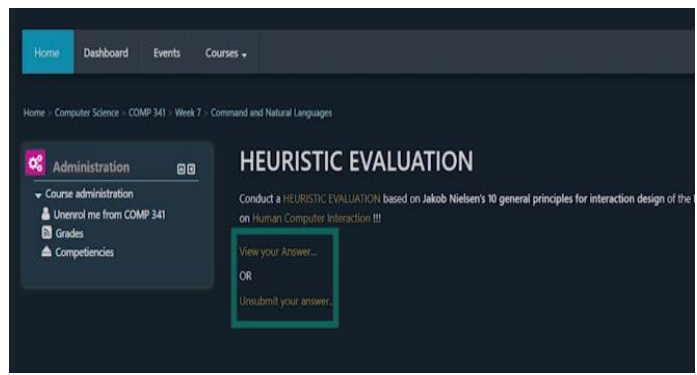

*Figure 2(a): Feature that allows to review and revise answers before submitting it. This feature serves as a safeguard against accidental submissions.*

*2.6.3.6 Recognition rather than recall*

Implementation of this principle is clearly observed in MOOC. Users can clearly see the remaining tasks in To Do List with the due date in the dashboard. From this section, users can see the remaining tasks and will be able to complete the task without much effort to remember and search about it as seen in figure 2(b) things to do are shown.

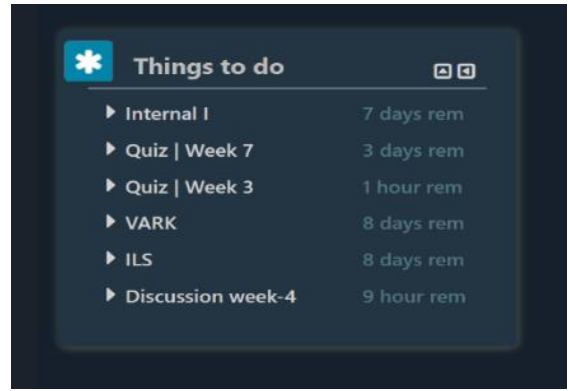

*Figure 1(b): Things to Do" feature helps stay organized. It is positioned in the top left corner so that it is visible for the user so that the user does not miss any important items on user's to-do list*

### 2.6.3.7 Flexibility and efficiency of use

The MOOC interface is flexible to use. For example, the tick marks in the weekly bar assures the completion of whole week work. This feature allows users to navigate quickly through the MOOC home page efficiently. Similarly, the accordion system adds efficiency to use the interface, as users can move from one week to another without much scrolling.

### 2.6.3.8 Aesthetic and minimalist design

The site is very good and has a minimalistic design. With the minimalistic design MOOC achieves the required task very easily. The dropdown accordion available in the new MOOC helps users track the course contents easily. Users do not need to scroll all the way down to see the tasks of week 1 or the previous lectures. The figure 3, shows accordion feature helps save space and indicates the completion status of weekly tasks. It prevents the need for extensive scrolling to find tasks.

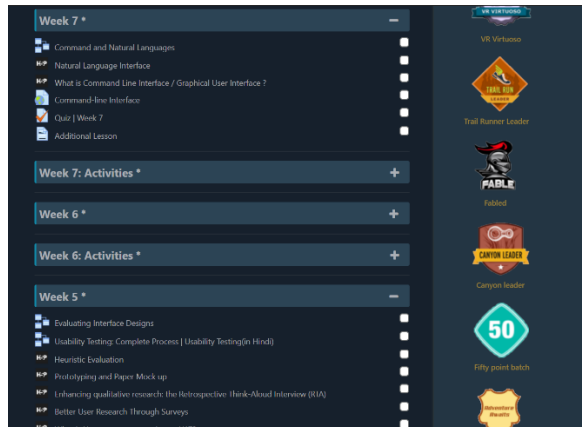

*Figure 2:The accordion feature*

*2.6.4 Acceptance Test and Evaluations during active use.*

After the design issues in methods were addressed by different types of evaluations, the prototype interface was made and tested, making users learn specific features, task performance (speed, error rate) and satisfactions with the pseudo inverted interface. Feedback from various groups, questionnaire and interview sessions were conducted during active use of the product by a targeted user which helped maintain the interface design. Interactions with MOOC users of different groups were made concerning respective interface designs for a evaluation and collecting feedbacks, so that all those can be addressed to develop a working prototype of interface.

## 3 RESULTS AND DISCUSSIONS

With the design principles of Human Computer Interaction specifically focused on the interaction of human in an online Learning platform, the purpose of our study was to find an existing problem, make modifications based on guidelines, design principles and evaluation criteria to the website to present an efficient modified website that resonates with user needs and better experience and helps them to stay within the page for a long time. With the evaluations, issues and modifications we were able to obtain the following results in different stages of our study:

### 3.1 Results from Survey

A survey was conducted using Google forms to distribute the prepared questionnaire and record the responses. A total of 107 responses were recorded for 31 questions each (one question being the user's Name). In our survey, students from different universities from more than 10 countries had participated. Most students in our survey were of age group 18 to 23 and 59.3% participants were male while 39.5% were female participants. 1.1% preferred not to say. The bar graph representation of survey result is shown in figure 4.

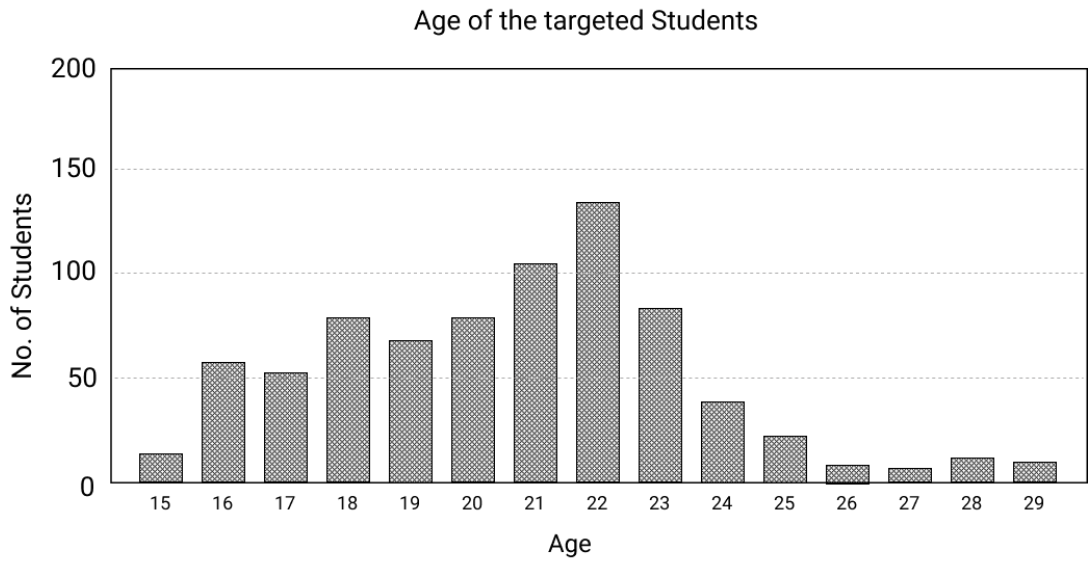

*Figure 4: Survey Results: Age Group of Students*

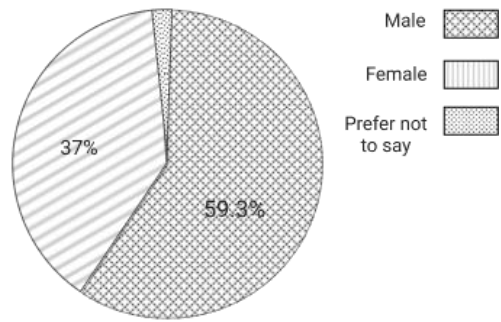

*Figure 5: Survey Results: Gender*

79.7% of the students responded that they prefer dark mode in their phone and 61.7% students said that they would love dark mode in learning sites as seen in figure 5. 73.7% students responded that it was more comfortable to read white text on a dark background which clearly showed that students do use and prefer dark mode in different applications and since our case was justified, we move ahead with our project.

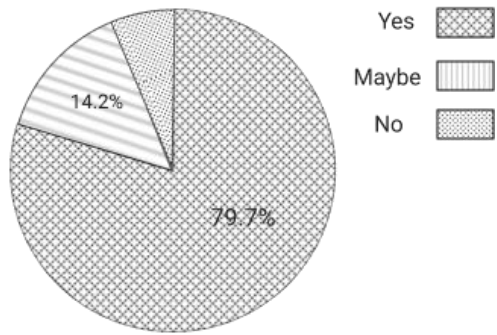

*Figure 6: Survey Results: Dark Mode Preferences*

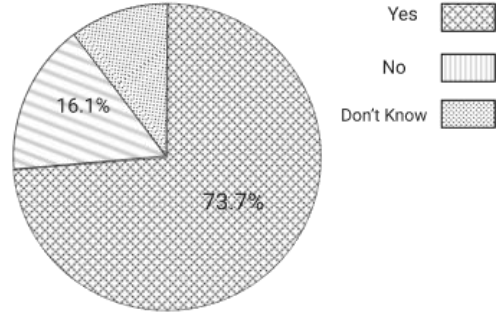

*Figure 7: Survey Results: Dark Mode Comfort Level*

Also 61.7% of the targeted students prefer dark mode in Learning Platforms as shown in figure 6 and figure 7. Coursera was taken as a reference so that higher accuracy could be achieved in the survey while 21.8% of the users denied the use of the dark mode in the learning platform making us to reach to the conclusion to provide an option to switch between the modes. Figure 8, shows dark mode preference in learning sites.

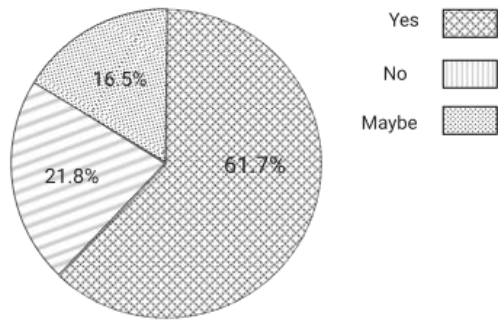

*Figure 8: Survey Results: Dark Mode Preference in Learning Sites*

## 3.2 Results from Evaluation of Design issues:

### 3.2.1 Results from Paper Prototyping.

On testing our initial paper prototype with 120 MOOC(Massive Open Online Course) users many of them found that their problems were issued in the prototype. We created a simple paper prototype of our case. Here, we took a short review of some experts and Mr. Chris R Becker in developing the paper prototype. During our survey we found that 61.90% of participants were totally satisfied by the new paper prototype. We also changed the color combinations analyzing the feedback of the users. The To DO section available in the new Dashboard was appreciated by the users and also some users suggested keeping the date/time in the To Do section. The icons for collapse and relapse of accordion of the week box were also changed following suggestions from MOOC users. The figure 9, shows results from users on paper prototype.

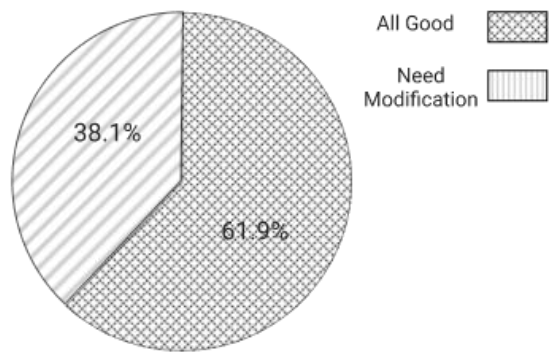

*Figure 9: Results from users on Paper Prototype*

*3.2.2* **Results from application of design principles:**

After conducting a survey from students, it was found that eLearning sites with no dark mode option was difficult to view for a longer period of time by the targeted users. edX which is a e-Learning site has dark mode option and it was found to be more comfortable and readable by users during night time. Jakob Nielsen's 10 general principles for interaction design was very effective in evaluating the eLearning websites and updating changes in them. Cognitive walkthrough was the best usability testing in this period. Expert Review from top UI/UX designers resulted in filtering out the most important missing elements of the existing websites to make modifications in the prototype. The figure 10, 11 shows the UI and features for the design. In the figure 13, Week 1 activities have been marked as completed so there is a tick marks present in top of accordion.

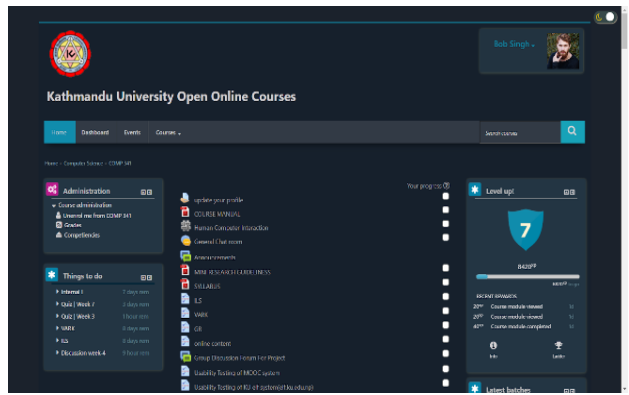

*Figure 10: Dark Mode User Interface designed after feedback from users.*

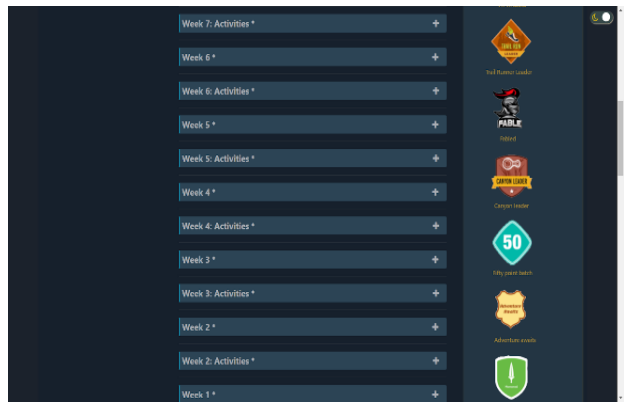

*Figure 11: Closed accordions save space and enhance visibility.*

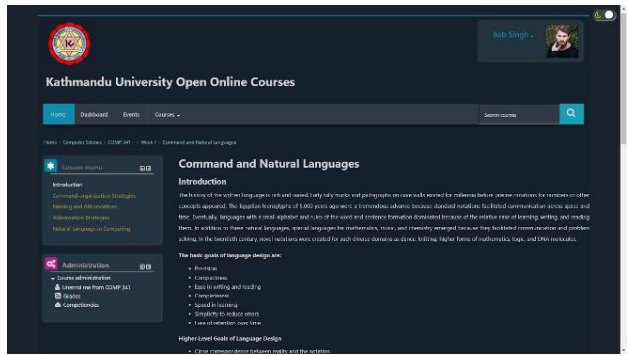

*Figure 12: Consistent theme for all pages*

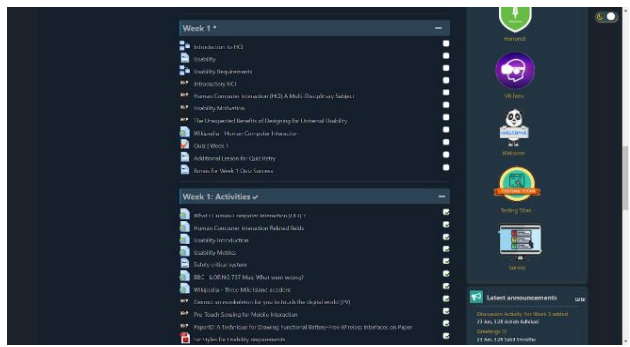

*Figure 13: The accordion that display completed week tasks*

### *3.2.3* **Results from usability testing.**

For usability testing, we chose some of the active users of MOOC and let our users use our system for studying during different intervals of time. 79% of the users strongly agreed that dark mode helped to reduce the eye strain and they would use this mode frequently while 72% users said that they would use the web application for a longer period of time. The figure 14, 15 shows the users feedback on use of dark modes like effects caused in eyes strain and dark mode usage period.

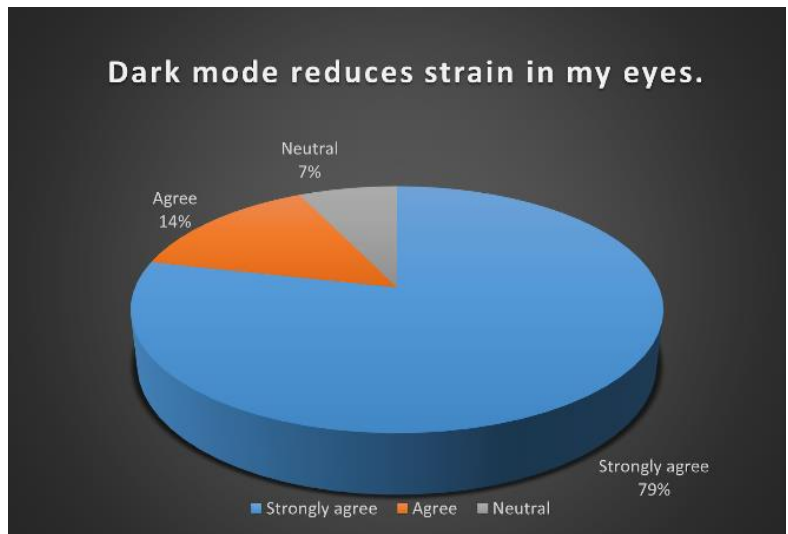
Figure 14. Dark mode strain reduction feedback

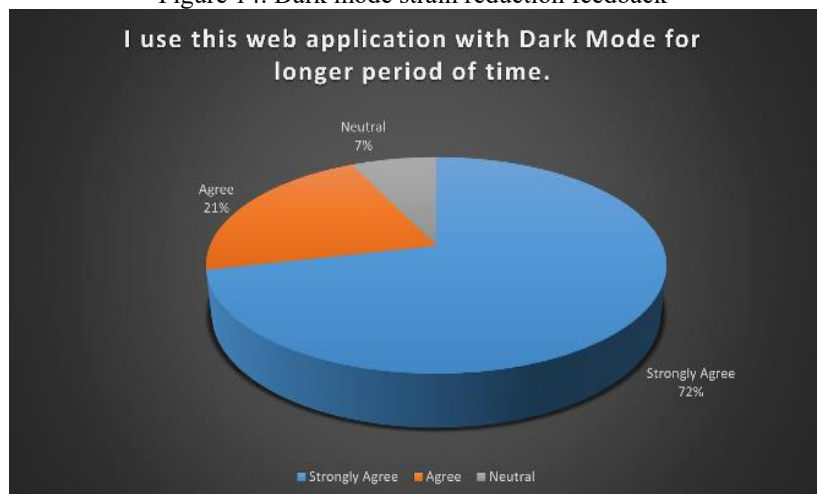
Figure 15. Longer period use feedbacks obtained

And, 71% users said that they preferred the dark mode in the MOOC. Thus, we found out that our dark theme was very much effective in reducing eye strain while studying for longer periods and motivated the users to study for longer times.

*3.2.4* **Results from student's survey.**

**Table 1:** Representation of Sample no., Mean and Standard deviation between groups showing different perspectives of using dark mode

| Descriptive Statistics | | | | | |
| --- | --- | --- | --- | --- | --- |
| **Sample** | **N** | **Minimum** | **Maximum** | **Mean** | **Std. Deviation** |
| Dark Mode reduces strain in my eyes | 707 | 1 | 5 | 3.26 | 1.233 |
| Dark themes are not always better for eye strain In bright light conditions | 707 | 1 | 5 | 3.05 | 1.081 |
| I think I will use the system with dark mode more frequently | 707 | 1 | 5 | 3.61 | 1.369 |
| I prefer to use dark mode in the study or learning platform I use if provided | 707 | 1 | 5 | 3.57 | 1.404 |
| I can study look at screen for study for longer time when I am using dark mode | 707 | 1 | 5 | 3.47 | 1.373 |
| Valid N (listwise) | 707 | | | | |

This table 1 shows a comprehensive set of descriptive statistics, offering insight into respondents' perceptions and attitudes towards the utilization of dark mode. Each statement is scrutinized through two distinct lenses: one employs a conventional 1 to 5 rating scale. In this survey, participants' opinions on dark mode were examined across five different aspects. On average, respondents generally agreed that dark mode reduces eye strain (mean = 3.26), although there was some variation in their responses (standard deviation = 1.233). However, they expressed a slightly lower level of agreement regarding the effectiveness of dark mode in bright light conditions (mean = 3.05) with relatively less variation (standard deviation = 1.081). Furthermore, participants indicated a moderate intention to use systems with dark mode more frequently (mean = 3.61), with some variability in their responses (standard deviation = 1.369). Similarly, they also exhibited a moderate preference for using dark mode in study or learning platforms when provided (mean = 3.57) with relatively higher variation (standard deviation = 1.404). Lastly, respondents reported a moderate level of agreement that they can study for longer periods while using dark mode (mean = 3.47), with some variation in their responses (standard deviation = 1.373). The valid number of responses considered for this analysis was 707, after removing any missing data.

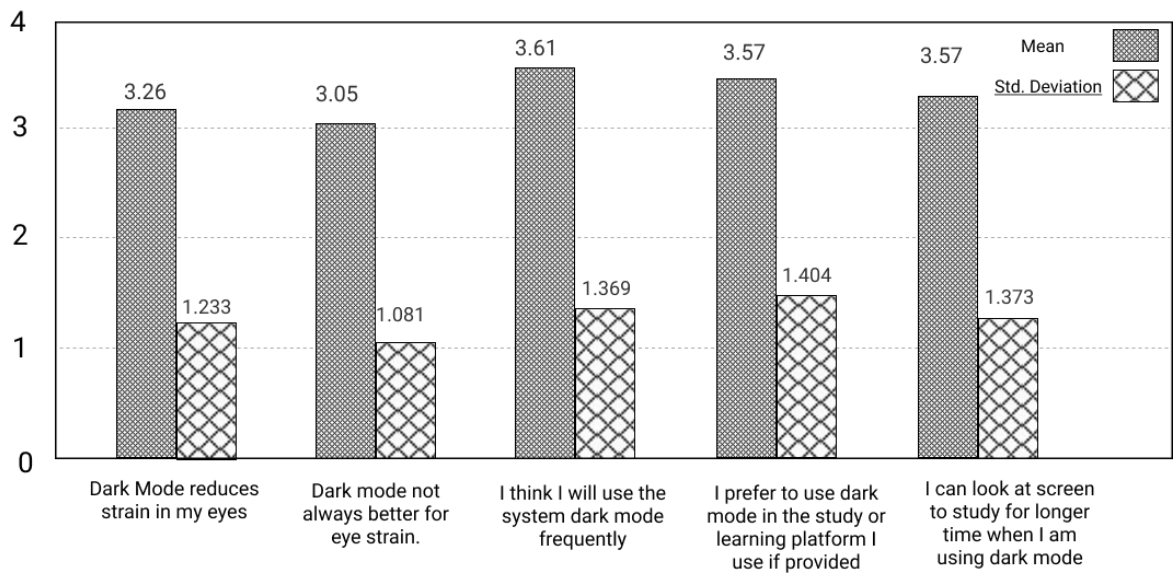

*Figure 16: Illustration of comparison between different perspectives of groups using dark mode*

Comparing mean scores as seen in figure 16, respondents generally agreed that dark mode reduces eye strain (mean = 3.26) but were slightly less convinced about its effectiveness in bright light conditions (mean = 3.05). They expressed strong intentions to use dark mode more frequently (mean = 3.61) and believed it could extend their study time (mean = 3.47). The standard deviations reflect varying degrees of consensus in these responses.

The data collected from our study gives a comprehensive understanding of student's preference and interactions with dark mode, focusing in the context of e-Learning sites. In this section we discuss the findings of our study about the study of students in relation to dark mode and its implications on Human Computer Interaction (HCI). We will discuss about each research question in relation to our findings.

**RQ1: Are there any subjective or objective benefits of using dark mode with respect to visual fatigue?**

According to the results of our poll, a sizable portion of students believe that dark mode lessens visual tiredness, as shown by the mean score of 3.26. This is in line with the body of research demonstrating that utilizing dark mode to limit blue light emissions from screens may have health benefits, including lessened eye strain. Our research indicates that dark mode should be included as a crucial feature for e-Learning platforms that value user experience.

**RQ2: Do students subjectively prefer dark mode or normal mode in their daily use cases?**

The research suggests a noticeable preference for dark mode amongst students, with 79.7% of the survey respondents favouring dark mode on their phones and 61.7% having interest in the feature for e-Learning sites. This does not imply the universal preference for dark mode, but it is an indication that a big section of the user base has an affinity to this mode. This sentiment also highlights the understanding of modern-day users, who are seeking more customized digital experiences focusing the visual comfort.

**RQ3: Do students tend to study for a longer period of time in the e-learning site with dark mode?**

The results indicate an inclination to dark mode when students tend to study for extended periods. A mean score of 3.47 suggests that survey respondents believe that they tend to study for longer hours with this mode. Meanwhile, 73.3% mentioning that they believe dark mode reduces visual fatigue supports that students believe in dark mode being easy for eyes. And the changes in HCI factor are also important here. Our questionnaires with MOOC users suggested that with little simple shifts in the aesthetics, it could lead to prolonged engagement in e learning sites, improving user retention and course completion rates.

Furthermore, the investigations and changes made to the MOOC system revealed the design challenges and resolutions that are necessary to make while converting a light mode system to dark mode. Incorporating the dark mode is not merely the exercise of aesthetics but also the user centred design to address the genuine issues. The designs, specially those that enhance the visual comfort can better the usability and functionality of the e learning sites.

While our study suggests the benefits of dark mode in e-learning sites, there may be some limitations to some aspects. Our target sample was mostly university students who are accustomed with modern UI designs and aware about latest changes and availability of options, which might not capture the preferences of older learners and those in non-academic settings. Moreover, objective measures such as eye tracking or physiological measures could provide deeper insights on the actual visual fatigue impacts.

## 4  FUTURE SCOPES

Future research on the implementation of dark mode in e-learning sites could delve into several promising areas. One potential direction is to conduct long-term studies to measure the sustained impact of dark mode on students' eye health, sleep patterns, and overall well-being. Additionally, exploring the psychological and cognitive effects of dark mode, such as its influence on concentration, information retention, and study efficiency, would provide valuable insights.

Examining how dark mode settings can be altered to accommodate personal tastes and visual comfort—such as changing the contrast and color schemes—is another option. To guarantee inclusivity and accessibility, research might also evaluate how dark mode affects other demographics, including age groups, those with visual impairments, and people with particular learning challenges. Ultimately, investigating how to include dark mode into e-learning platforms that use cutting-edge technologies like augmented reality (AR) and virtual reality (VR) could improve the immersive learning environment while reducing the possible negative effects of extended screen time.

## 5 CONCLUSION

E-learning platforms are very popular now, and researchers and developers are working to enhance user experience with their systems. Our research, considering the effects of viewing a brighter background for longer hours in a darker environment, helps to identify the difficulty that students face during the time of interacting with various e-Learning sites especially during late hours. The user preferences and behaviours are analysed using different techniques of HCI like survey, interview, evaluation methods and principles of design.

From our research, it is clear that a significant number of students believe that dark mode reduces visual fatigue and makes it easier for them to concentrate and comprehend text in low-light conditions. Also, many students Favor using dark mode on their devices on a regular basis, and on e-learning platform as well. This suggests not only a rising trend, but an echoing sentiment that points to shift towards the interfaces with visual and health comfort. As the digital landscape continues to evolve, paying attention to user preferences and developing customized user centred designs, especially those with better visual comforts can prolong the user engagement, reduced visual strain and enhanced experience. In this way the dark mode plays a vital role in HCI of students in e-Learning sites.